\documentclass[11pt,letterpaper]{article}
\usepackage[utf8]{inputenc}
\usepackage{multirow} 
\usepackage{booktabs} 
\usepackage[USenglish,spanish]{babel}
\usepackage{amsmath}
\usepackage{amsfonts}
\usepackage{amssymb}
\usepackage{graphicx}
\usepackage[left=3cm,right=3cm,top=2cm,bottom=3cm]{geometry}

\usepackage[small,bf]{caption} 

\author{Jorge Pinochet}
\title{\textbf{La desigualdad de Bell y las fantasmagóricas acciones a distancia \\ 
\large Bell's inequality and the spooky actions at a distance}}

\begin{document}

\renewcommand{\figurename}{\textbf{Figura}}

\author{Jorge Pinochet$^{*}$\\ \\
 \small{$^{*}$\textit{Facultad de Ciencias Básicas, Departamento de Física. }}\\
  \small{\textit{Centro de Investigación en Educación (CIE-UMCE),}}\\
 \small{\textit{Núcleo Pensamiento Computacional y Educación para el Desarrollo Sostenible (NuCES).}}\\
 \small{\textit{Universidad Metropolitana de Ciencias de la Educación,}}\\
 \small{\textit{Av. José Pedro Alessandri 774, Ñuñoa, Santiago, Chile.}}\\
 \small{e-mail: jorge.pinochet@umce.cl}\\}

\date{} 
\maketitle

\begin{center}\rule{0.9\textwidth}{0.1mm} \end{center}
\begin{abstract}
\noindent En este artículo se discuten los principales aspectos relacionados con la desigualdad de Bell, tanto teóricos como experimentales. También se presenta una nueva derivación de la desigualdad de Bell, que destaca por su simplicidad matemática. La exposición está destinada principalmente a estudiantes no graduados de física, y pone especial énfasis en clarificar el significado y los alcances de la desigualdad de Bell en el contexto del experimento de Einstein-Podolski-Rosen.\\

\noindent \textbf{Descriptores}: Desigualdad de Bell, experimento EPR, entrelazamiento cuántico, desigualdad de Wigner-d´Espagnat.   

\end{abstract}

\selectlanguage{USenglish}

\begin{abstract}
\noindent This article discusses the main aspects related to Bell's inequality, both theoretical and experimental. A new derivation of Bell's inequality is also presented, which stands out for its mathematical simplicity. The exposition is mainly intended for undergraduate physics students, and places special emphasis on clarifying the meaning and scope of Bell's theorem in the context of the Einstein-Podolski-Rosen experiment. \\


\begin{center}\rule{0.9\textwidth}{0.1mm} \end{center}
\end{abstract}

\selectlanguage{spanish}

\maketitle

\renewcommand{\tablename}{Tabla} 

\section{Introducción}

La mayoría de las personas dan por sentado que dos o más objetos separados espacialmente, no pueden afectarse mutuamente si están lo suficientemente alejados unos de otros. Esta idea, que los físicos denominan \textit{principio de localidad} o simplemente \textit{localidad}, durante mucho tiempo fue considerado un dogma sagrado de la física. Uno de los más férreos defensores de este dogma fue Albert Einstein, quien calificaba cualquier posible violación de la localidad como una “fantasmagórica acción a distancia” [1].\\

Einstein utilizó el dogma de la localidad para intentar demostrar que la descripción de la realidad física proporcionada por la mecánica cuántica (MC) es incompleta. Para llevar a cabo la demostración, Einstein elaboró un ingenioso experimento mental junto a los físicos Boris Podolsky y Nathan Rosen, que hoy en día conocemos como \textit{experimento EPR}, en honor sus autores [2]. El experimento mental involucra dos partículas que interactúan durante un breve lapso y luego se separan\footnote{En la versión original del experimento EPR, Einstein y sus colaboradores no hablan de dos partículas sino de dos sistemas. No obstante, las versiones modernas del experimento que se encuentran en artículos y textos de mecánica cuántica se referieren a pares de partículas.}. Asumiendo que la localidad no puede violarse, el experimento EPR concluye que la descripción cuántica de las partículas no contiene toda la información física pertinente, lo que implica que la MC es incompleta. Según Einstein, esta incompletitud es la responsable de que la MC sea una teoría probabilística, pues, al trabajar con información truncada, es incapaz de suministrar una descripción determinista de los fenómenos físicos.\\

Una importante limitación del experimento EPR, al menos en la forma que Einstein y sus colaboradores lo concibieron en 1935, es que resultaba imposible convertirlo en un experimento real. En la práctica, esto significaba que el principio de localidad quedaba fuera del ámbito de lo empíricamente comprobable. Casi 30 años después, en 1964, el físico teórico irlandés John S. Bell generalizó el experimento EPR a un escenario donde hay múltiples pares de partículas [3]. Bell demostró que, en este nuevo escenario, cabe esperar dos tipos de correlaciones estadísticas entre los pares de partículas: (1) correlaciones clásicas, donde los efectos son locales (las partículas no pueden afectarse mutuamente si están muy alejadas una de otra); (2) correlaciones cuánticas, donde los efectos son no locales (las partículas pueden afectarse mutuamente aunque se encuentren a gran distancia una de otra). Bell también derivo matemáticamente un resultado conocido como \textit{teorema de Bell} o \textit{desigualdad de Bell}, que resulta satisfecho por el primer tipo de correlaciones, pero que es violado por el segundo. Este importante hallazgo abrió las puertas a la contrastación empírica del experimento EPR y del principio de localidad.\\ 

A partir de la década de 1970 distintos grupos de investigadores comenzaron a diseñar y ejecutar experimentos destinados a verificar una posible violación del teorema de Bell [4–6]. En los años transcurridos desde entonces, los experimentos han mostrado de forma concluyente que el teorema es violado. Si adoptamos un punto de vista realista, es decir, si suponemos que los objetos que conforman el mundo físico tienen una existencia objetiva, independiente del sujeto que lo observa, la violación del teorema de Bell implica que los efectos no locales forman parte del mundo microscópico, y que, por lo tanto, las fantasmagóricas acciones a distancia que Einstein rechazaba son reales. Hoy en día preferimos utilizar un nombre menos fantasmagórico para esta violación de la localidad, llamándola \textit{entrelazamiento cuántico}, un término acuñado por Erwin Schrödinger que describe muy bien la esencia de este fenómeno, que constituye una de las propiedades más sorprendentes y fundamentales de la MC. Un conjunto de partículas entrelazadas no pueden definirse como entidades independientes, sino que deben describirse como un sistema cuyos componentes están conectados mediante interacciones superluminales. El entrelazamiento cuántico no es sinónimo de no localidad, pero si adoptamos un punto de vista realista, podemos considerar que ambos términos son equivalentes. \\

Entre los físicos experimentales que pusieron a prueba la desigualdad de Bell, destaca el trabajo pionero del estadounidense John Clauser, del francés Alain Aspect, y del austriaco Anton Zeilinger, quienes fueron galardonados con el Premio Nobel de Física 2022 por su importante trabajo en este campo [7–9]. Este acontecimiento científico abre la oportunidad para que especialistas y educadores del área intentemos explicar la desigualdad de Bell a un público lo más amplio posible, ya que, salvo por un puñado de especialistas, el trabajo de Bell es desconocido por la inmensa mayoría de la gente, incluyendo a muchos físicos profesionales. \\

Durante las últimas décadas, diversos autores han asumido el desafío de simplificar y divulgar el trabajo de Bell [10–27]. De todas las iniciativas propuestas hasta ahora, la que mejor parece conjugar simplicidad y profundidad es un artículo debido a Bernard d´Espagnat [26], donde se introduce una versión elemental del teorema de Bell conocida como \textit{desigualdad de Wigner-d´Espagnat} (WD) [12,26]. El objetivo de este trabajo es doble: Por una parte, busca ofrecer a los no especialistas una exposición simple y actualizada sobre la desigualdad de Bell, y por otra, busca proporcionar una nueva derivación matemática de la desigualdad WD que destaca por su simplicidad matemática y conceptual. Un aspecto clave de la derivación es que se presenta como una extensión directa del experimento EPR. El artículo está dirigido a aquellos lectores que dominan el algebra elemental, y que están familiarizados con la MC.\\

El artículo comienza desarrollando una versión simplificada y no técnica del experimento EPR. A continuación, se introduce la derivación elemental de la desigualdad WD, presentándola como una extensión directa del experimento EPR. Luego se contrasta la desigualdad WD con las predicciones probabilísticas de la MC, tanto en términos numéricos como gráficos. A continuación, se analiza el significado físico de la violación de la desigualdad de Bell en el contexto de los resultados experimentales. Después, se contrastan las visiones filosóficas del realismo y del positivismo como marcos interpretativos de la violación experimental de la desigualdad de Bell. El artículo finaliza con unas breves reflexiones.\\ 

Antes de abocarnos a nuestra tarea, es importante advertir al lector que el tema que abordaremos no solo tiene profundas implicancias físicas sino también filosóficas. Es justamente en el terreno de la filosofía donde el teorema de Bell y el principio de localidad suscitan más debate y desacuerdo entre los expertos. La razón es simple: la filosofía se mueve en el terreno de las interpretaciones y no de los resultados experimentales, que suelen concitar amplios consensos debido a su mayor grado de objetividad. Como este artículo está dirigido a los no especialistas, nos interesa evitar, dentro de lo posible, las complejas y áridas discusiones filosóficas que seguramente distraerían al lector de nuestro objetivo principal. Como esto nos obliga a tomar una posición filosófica desde un comienzo, adoptaremos un punto de vista realista. El realismo fue justamente el enfoque adoptado por Einstein y Bell, y es el enfoque quizá más provechoso con fines explicativos. No obstante, antes de finalizar el artículo mostraremos al lector que el realismo no es la única postura posible.

\section{El experimento EPR}

El trabajo original de Bell se basó en una versión simplifica del experimento EPR, ideada por el físico David Bohm en 1951 [28]. Al igual que en la versión de 1935, el experimento mental de Bohm, o \textit{experimento EPRB} como lo conocemos hoy en día, tenía como objetivo demostrar que la MC es una teoría incompleta, lo que significa que su formalismo matemático no contiene toda la información sobre los sistemas físicos bajo estudio. Como el teorema de Bell es un resultado estadístico que establece correlaciones entre un gran número de medidas, nuestro objetivo en esta sección es avanzar hacia una generalización del trabajo Bohm, que considera los resultados obtenidos al realizar múltiples experimentos EPRB de forma secuencial. Pero antes de ello debemos entender la lógica que subyace al experimento individual.\\

El experimento EPRB considera un par de protones que son emitidos desde una fuente en direcciones opuestas hacia dispositivos especializados que miden sus componentes de spin en dos direcciones mutuamente perpendiculares, $z$ y $x$ [28]. Llamemos Alice al dispositivo que está a la izquierda de la fuente, y Bob al dispositivo que está a la derecha. Alice y Bob pueden rotarse para medir el spin ya sea en la dirección $z$ o en la $x$. Las distancias de Alice y Bob a la fuente son iguales. Luego de ser emitidos, los protones se separan viajando en sentidos opuestos a lo largo de un mismo eje, sin sufrir ninguna perturbación. Como quedará claro enseguida, esta condición de no perturbación es muy importante, pues, de no cumplirse, no podríamos desarrollar la argumentación en que se basa el experimento EPRB.\\

Conceptualmente, el spin es una de las propiedades cuánticas más simples, y es el equivalente cuántico del momentum angular clásico, es decir, es un vector que mide la rotación intrínseca de una partícula. Pero hay que tener cuidado con las analogías, ya que, en el marco de la Física Clásica, el momento angular en una determinada dirección puede tener un continuo de resultados entre dos valores extremos, mientras que, de acuerdo con la MC y con los resultados experimentales, el spin puede tomar solo dos valores en una dirección dada, denotados convencionalmente como $+$ y $-$, de manera que se trata de una propiedad dicotómica [25]. Los posibles valores de una mediada de spin en la dirección $z$ se designan como $z+$ y $z-$, y los posibles valores en la dirección $x$ se designan como $x+$ y $x-$. Vale decir, si Alice y/o Bob están orientados para medir en la dirección $z$, solo podrán encontrar los valores $z+$ o $z-$, y si están orientados en la dirección $x$, solo podrán encontrar los valores $x+$ o $x-$.\\

En la jerga técnica de la MC se dice que las componentes de spin en diferentes direcciones son \textit{observables incompatibles}, lo que significa que dichas componentes no pueden medirse simultáneamente sobre una misma partícula. En el experimento EPRB los pares de protones detectados por Alice y Bob se encuentran en un estado denominado \textit{singlete}, que es el equivalente cuántico del estado clásico de momentum angular nulo [29]. Por conservación del momentum angular esto significa que, si el spin de un protón es $+$ en una dirección dada, el del otro es $-$ en la misma dirección, y viceversa, de manera que, vectorialmente, el spin total de ambas partículas es cero. Si suponemos que Alice y Bob se encuentran orientados para medir el spin en la dirección $z$, el estado singlete en que se encuentran los pares de protones del experimento EPRB se escribe formalmente como [29]:

\begin{equation}
\vert \psi \rangle =\frac{1}{\sqrt{2}} \vert z+ \rangle_{A} \vert z- \rangle_{B} - \frac{1}{\sqrt{2}}\vert z- \rangle_{A} \vert z+ \rangle_{B}.
\end{equation}

Esta fórmula nos dice que los protones están en un estado compuesto por dos alternativas mutuamente excluyentes que describen las probabilidades de una medida conjunta de spin en la dirección $z$. Según la Ec. (1), el caso en que la componente de spin medida por Alice ($A$) es $z+$ y la componente medida por Bob ($B$) es $z-$ se representa como $\vert z+ \rangle_{A} \vert z- \rangle_{B}$. De acuerdo con la regla de Born, la probabilidad de que esto ocurra se calcula elevando al cuadrado el coeficiente numérico correspondiente [30], en este caso $1/\sqrt{2}$, de manera que la probabilidad es $1/2$. Por otra parte, el caso en que la componente medida por Alice es $z-$, y la componente medida por Bob es $z+$, se representa como $\vert z- \rangle_{A} \vert z+ \rangle_{B}$, donde nuevamente la probabilidad de que esto ocurra es $1/2$. Notemos que estas probabilidades cumplen la condición de normalización, es decir, cubren todas las posibilidades, de manera que, al sumarlas, se obtiene la unidad ($1/2 + 1/2 = 1$), como cabe esperar si las probabilidades están bien definidas. Como la orientación de los ejes de medición de Alice y Bob es arbitraria, las conclusiones anteriores se mantienen si los ejes coinciden con la dirección $x$.\\

La importancia del estado singlete para el desarrollo del experimento EPRB reside en la siguiente propiedad fundamental, que fue mencionada más atrás: por conservación del momentum angular, cuando Alice y Bob efectúan medidas conjuntas de spin sobre un mismo eje, la MC predice resultados opuestos para dichas medidas. En otras palabras, la Ec. (1) predice una correlación negativa perfecta entre medidas de spin efectuadas sobre un mismo eje. Qué valor tendrá la componente en una dirección determinada es completamente impredecible; sin embargo, la Ec. (1) nos garantiza que, si Alice mide $+$, Bob medirá $-$, y viceversa. \\

De acuerdo con la lógica del experimento EPRB, imaginemos que Alice mide el espín en la dirección $z$ y Bob en la dirección $x$. Por las propiedades del estado singlete se sigue que si Alice mide $z+$, entonces podemos inferir con certeza que, si Bob midiera sobre el eje $z$, el resultado sería $z-$. Como se trata de una medida conjunta de spin, al mismo tiempo que Alice mide $z+$, podemos imaginar que Bob mide $x+$, lo que significa que, si Alice midiera sobre el eje $x$, con certeza el resultado sería $x-$ [25]. Así, al protón detectado por Alice le asignamos una componente de spin medida $z+$, y al protón detectado por Bob le asignamos una componente inferida $z-$. De igual modo, al protón detectado por Bob le asignamos una componente medida $x+$, y al protón detectado por Alice le asignamos una componente inferida $x-$. Pero también podríamos haber llegado a esta conclusión invirtiendo los ejes de medición, de modo que Alice mida en la dirección $x$ y Bob en la dirección $z$. Como existen dos ejes de medición, $x$ y $z$, y dos valores de spin para cada eje, $+$ y $-$, se sigue que hay $2^{2}=4$ resultados posibles para las medidas e inferencias efectuadas por Alice y Bob sobre un par de protones. Estos cuatro resultados se resumen en la Tabla 1.\\

\begin{table}[htbp]
\begin{center}
\caption{Medidas e inferencias de componentes de spin en las direcciones $z$ y $x$.}
\begin{tabular}{l l l l} 
\toprule
Medida de Alice & Inferencia de Alice & Medida de Bob & Inferencia de Bob\\
\toprule
$z+$ & $z-$ & $x-$ & $x+$ \\ 
\midrule
$z-$ & $z+$ & $x+$ & $x-$ \\ 
\midrule
$x+$ & $x-$ & $z+$ & $z-$ \\
\midrule
$x-$ & $x+$ & $z-$ & $z+$ \\
\bottomrule
\end{tabular}
\label{Medidas e inferencias de componentes de spin en las direcciones z y x}
\end{center}
\end{table}

Es muy importante tener presentes los dos supuestos en que se basa el experimento EPRB. En relación con el primer supuesto se observa, por ejemplo, que al inferir un resultado $z-$ de Bob a partir de una medida $z+$ de Alice, estamos asumiendo que la componente de spin $z-$ tiene existencia real, con independencia del proceso de medición. En otras palabras, ya sea que midamos o no la componente $z-$, el hecho de poder predecir con certeza su valor nos permite afirmar que dicha componente es una propiedad real y objetiva del protón. Este primer supuesto, que ya mencionamos en la introducción, se conoce como \textit{realismo}, y sostiene que existe una realidad objetiva que es independiente del observador. En relación con el segundo supuesto, como de acuerdo con la física relativista, ninguna señal o interacción puede propagarse más rápido que la luz, si al momento de realizar las medidas conjuntas de spin, Alice y Bob se encuentran lo suficientemente alejados uno de otro, podemos descartar la existencia de señales o interacciones que se propaguen entre los protones o entre los dispositivos de medición, que alteren las correlaciones negativas de la Tabla 1. Este segundo supuesto es la \textit{localidad}, y como sabemos, establece que dos o más objetos separados espacialmente, no pueden afectarse mutuamente si están lo suficientemente alejados unos de otros. Expresado de forma más precisa, la localidad nos dice que dos o más objetos separados una distancia arbitrariamente grande no pueden interactuar físicamente mediante influencias cuya rapidez de propagación sea superior a la de la luz. \\

En conjunto, el realismo y la localidad conducen a lo que se ha dado en llamar \textit{realismo local} [26]. El experimento EPRB se basa en la hipótesis de que el realismo local es correcto. Bajo estas condiciones, las medidas e inferencias de Alice y Bob están en pie de igualdad, púes ambas describen propiedades objetivas de las partículas. Esto nos permite elaborar una nueva tabla, la Tabla 2, que resume los resultados de la Tabla 1, pero no distingue entre medidas e inferencias, y por lo tanto asume el realismo local como hipótesis fundamental. Esta tabla también incorpora una columna adicional llamada “cardinalidad”, cuyo significado quedará claro un poco más adelante.\\ 

\begin{table}[htbp]
\begin{center}
\caption{Componentes de spin en las direcciones $z$ y $x$.}
\begin{tabular}{l l l} 
\toprule
Cardinalidad & Alice & Bob \\
\toprule
$n_{1}$ & $z+,x+$ & $z-,x-$ \\ 
\midrule
$n_{2}$ & $z+,x-$ & $z-,x+$ \\ 
\midrule
$n_{3}$ & $z-,x+$ & $z+,x-$ \\
\midrule
$n_{4}$ & $z-,x-$ & $z+,x+$\\
\bottomrule
\end{tabular}
\label{Componentes de spin en las direcciones z y x}
\end{center}
\end{table}

De momento, lo que no interesa destacar de la Tabla 2 es que los resultados que muestra contradicen a la MC ya que, según las reglas cuánticas, las componentes de spin en las direcciones $z$ y $x$ son observables incompatibles, y por lo tanto, es imposible determinarlos simultáneamente sobre un mismo protón. Si suponemos que Alice mide en la dirección $z$ y Bob en la $x$, el único resultado compatible con la MC es que la componente de spin del protón detectado por Alice sea $z+$ o $z-$ y la componente del protón detectado por Bob sea $x+$ o $x-$; si suponemos que Alice mide en la dirección $x$ y Bob en la $z$, el único resultado compatible con la MC es que la componente de spin del protón detectado por Alice sea $x+$ o $x-$ y la componente del protón detectado por Bob sea $z+$ o $z-$. Como los resultados obtenidos por Alice y Bob, resumidos en la Tabla 2, incorporan más información de la autorizada por las reglas cuánticas, concluimos que la descripción suministrada por la MC es incompleta o aproximada [11]. Para evitar esta conclusión, tendríamos que rechazar el supuesto en que se basa el experimento EPRB, esto es, el realismo local\\

Para ilustrar el absurdo al que, a su juicio, conduce el rechazo del realismo local, Einstein se preguntaba: Is the moon there when nobody looks? [31]. El realismo local es una característica fundamental de la física clásica, lo que incluye a la teoría especial y general de la relatividad de Einstein, pero como veremos pronto, la MC no satisface el realismo local, y la desigualdad de Bell es la herramienta que nos permite demostrarlo. En efecto, de acuerdo con esta idea, y siguiendo el análisis efectuado originalmente por Bell, generalicemos el experimento EPRB a un escenario donde hay $N$ pares de protones en estado singlete que son emitidos secuencialmente desde la fuente. Alice y Bob miden el espín de cada par de protones, lo que equivale a realizar el experimento EPRB $N$ veces, donde suponemos que $N$ es grande.\\ 

Como sabemos, la situación que interesa analizar es cuando las medidas de Alice y Bob se efectúan sobre ejes mutuamente perpendiculares, tales como $z$ y $x$ (Fig. 1). En algunos momentos Alice mide el espín en la dirección $z$ y Bob en la dirección $x$, y en otros momentos Alice mide el espín en la dirección $x$ y Bob en la dirección $z$. Nuevamente, como existen dos ejes de medición y dos valores de spin para cada eje, $+$ y $-$, se sigue que hay $2^{2}=4$ resultados posibles. Estos resultados dan origen a los cuatro conjuntos que aparecen en la Tabla 2, que resumen todas las medidas de spin efectuadas por Alice y Bob sobre $N$ pares de protones en $N$ experimentos EPRB.\\

\begin{figure}[h]
  \centering
    \includegraphics[width=0.6\textwidth]{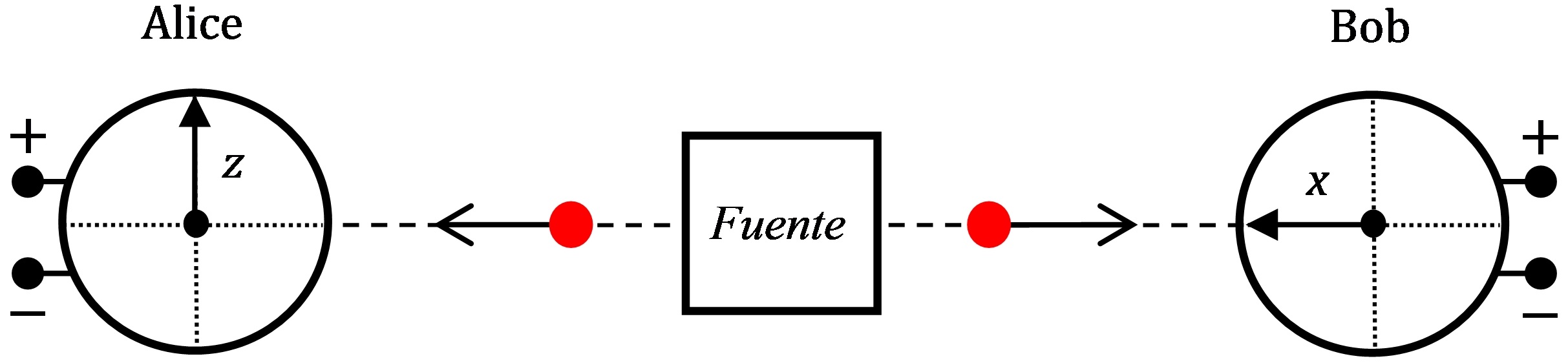}
  \caption{Dos protones en estado singlete (círculos rojos) viajan a lo largo de un mismo eje en sentidos opuestos. Un protón viaja hacia Alice, que mide el spin en la dirección $z$, y el otro viaja hacia Bob, que mide en la dirección $x$. Cuando Alice obtiene el resultado $+/–$, Bob obtiene el resultado $–/+$.}
\end{figure}

La Tabla 2 revela una correlación negativa perfecta entre las medidas. Cada conjunto tiene una cardinalidad (número de elementos), que corresponde a la cantidad de pares de protones en estado singlete con las componentes de spin indicadas, donde $N=n_{1}+n_{2}+n_{3}+n_{4}$. El conjunto con cardinalidad $n_{1}$ reúne todos los pares donde el spin del protón detectado por Alice tiene componentes $z+,x+$ y el spin del protón detectado por Bob tiene componentes $z-,x-$; el conjunto con cardinalidad $n_{2}$ reúne todos los pares donde el spin del protón detectado por Alice tiene componentes $z+,x-$ y el spin del protón detectado por Bob tiene componentes $z-,x+$, etc. Aunque la Tabla 1 considera las posibles medidas sobre un solo par de protones, mientras que la Tabla 2 considera un gran número de medidas sobre muchos pares de protones, ambas tablas contradicen a la MC ya que, según las reglas cuánticas, los componentes de espín en las direcciones $z$ y $x$ son observables incompatibles. Por lo tanto, en el marco del experimento EPRB, y de acuerdo con la hipótesis del realismo local, las $N$ medidas de spin resumidas en la Tabla 2 también nos dicen que la MC es incompleta o aproximada [11]. 

\section{Los calcetines de Bertlmann y el experimento EPRB}

Quizá el lector no esté muy impresionado con los resultados del experimento EPRB, y tampoco entienda por qué ha provocado tanto alboroto entre los físicos, ya que existen muchos ejemplos cotidianos que reproducen el tipo de correlaciones clásicas entre pares de protones que describe el experimento EPRB. Un divertido ejemplo es el caso del Dr. Bertlmann, un físico excéntrico al que le gusta llevar siempre calcetines de distinto color [10]. Qué colores llevará un día determinado es impredecible. Sin embargo, si un cierto día observamos que lleva un calcetín rosado en un pie dado, podemos tener certeza de que el calcetín del otro pie no será rosado. ¿Los pares de protones en estado singlete del experimento EPRB se comportan del mismo modo que los pares de calcetines de Bertlmann? \\ 

Según Einstein, la respuesta en un categórico sí. Al observar, por ejemplo, que el calcetín del pie izquierdo es rosado, podemos determinar con certeza y objetividad que el color del calcetín del pie derecho es no-rosado, con independencia de que lo hayamos observado. Según el experimento EPRB, las inferencias respecto del spin de los protones son tan válidas como las inferencias respecto del color de los calcetines de Bertlmann. Así, al observar que el protón izquierdo tiene spin $+$, podemos determinar con certeza y objetividad que el protón derecho tiene spin $-$, con independencia de que lo hayamos medido. Seguramente la mayoría de nosotros estaría de acuerdo con esta postura, que está en perfecto acuerdo con el realismo local. Sin embargo, la MC nos presenta un panorama muy distinto. \\

\begin{figure}[h]
  \centering
    \includegraphics[width=0.4\textwidth]{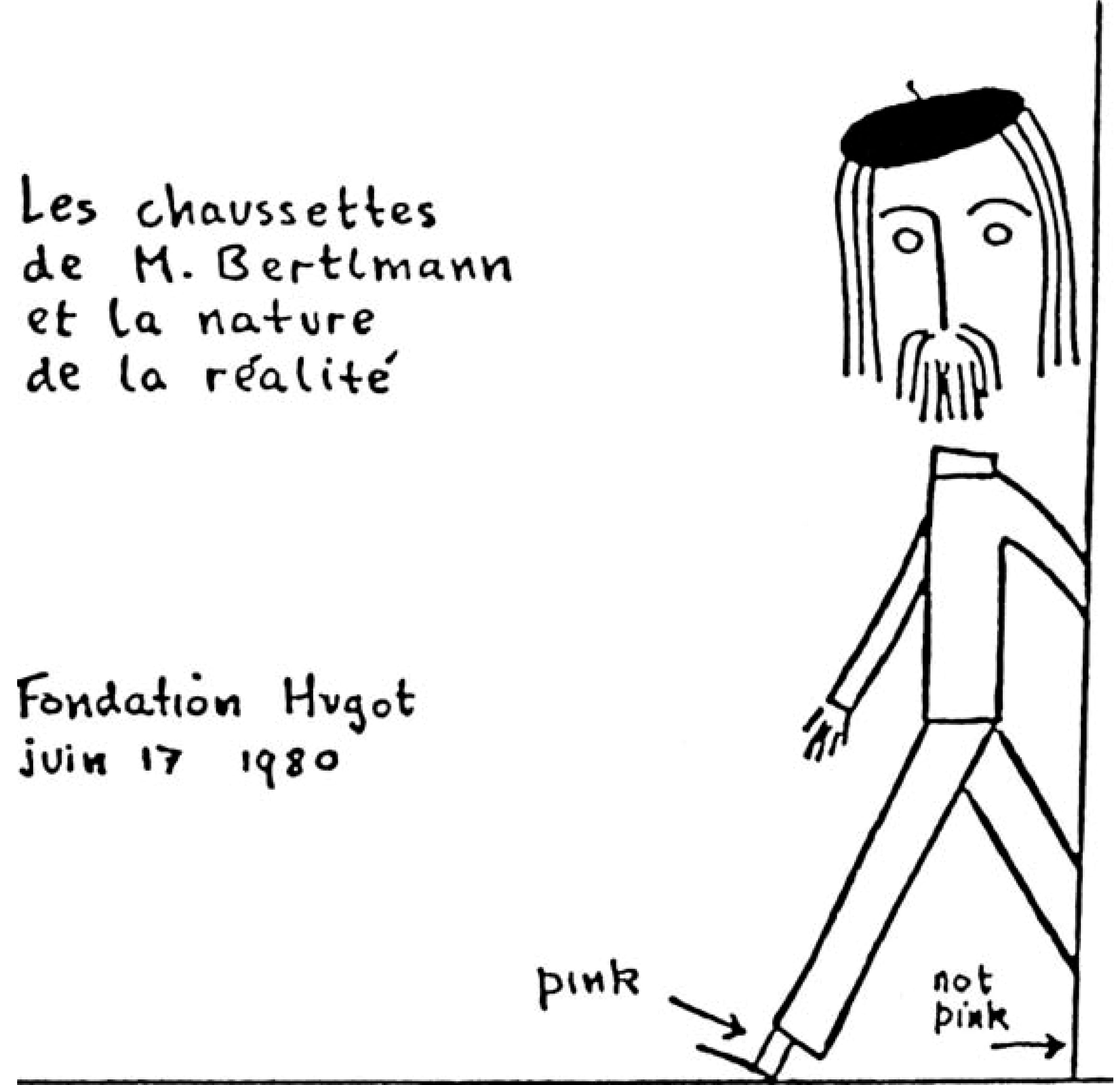}
  \caption{El caso de los calcetines de Bertlmann ilustra la diferencia entre las correlaciones clásicas del experimento EPRB y las correlaciones cuánticas (Fuente: https://physicstoday.scitation.org/na101/home/literatum/publisher/aip/journals/content/pto/2015/pto.2015.68.issue-7/pt.3.2847/production/images/large/pt.3.2847.figures.f2.jpeg).}
\end{figure}

Si los calcetines de Bertlmann fueran pares de protones en estado singlete, desde un punto de vista operacional, la MC nos dice que no podemos atribuirles un color definido hasta que los observamos, y una vez que vemos que el calcetín izquierdo es de un color determinado, el calcetín derecho adquiere automáticamente el color opuesto. Este escenario cuántico viola la localidad, pues asume una conexión instantánea a distancia entre los pares de calcetines, que se comportan como una solo entidad y no como dos objetos independientes. Esta conexión a distancia es lo que hoy en día denominamos entrelazamiento cuántico, un concepto que Einstein consideraba absurdo, y que calificaba como una fantasmagórica acción a distancia. Reformulemos la pregunta planteada antes a la luz de estas ideas: ¿Están entrelazados los pares de protones en estado singlete, o simplemente se comportan como los pares de calcetines de Bertlmann en una sola dirección de componente de espín? \\ 

La principal dificultad para responder a esta pregunta es que, en el marco del experimento EPRB, los resultados obtenidos al medir el spin de pares de protones entrelazados son indistinguibles de los resultados provenientes de una inferencia a lo Bertlmann. Empíricamente, los pares de protones entrelazados se comportan del mismo modo que los pares de calcetines. Como el experimento EPRB solo considera ejes de medición de spin cuyas orientaciones relativas son de $0^{o}$ y $90^{o}$, Bell decidió investigar lo que sucede en una situación más general, donde se mantiene la lógica de Bertlmann, pero los ejes no son colineales ni perpendiculares. Descubrió que, bajo estas nuevas condiciones, las correlaciones predichas por el escenario EPRB son solo un subconjunto del espectro más amplio de resultados que pueden obtenerse en el escenario cuántico [25]. Como veremos pronto, esta diferencia se pone de manifiesto a través del teorema Bell. Formalmente, el teorema es una desigualdad matemática que solo es satisfecha bajo el supuesto de que la hipótesis del realismo local es correcta. Como las predicciones de la MC no satisfacen la desigualdad de Bell, violan el realismo local, lo que significa que desde el punto de vista de la MC los pares de protones del experimento EPRB están entrelazados, y por lo tanto, interactúan mediante influencias que se propagan más rápido que la luz, con independencia de la distancia que los separe.\\  

Hasta aquí, lo que hizo Bell fue demostrar que el realismo local es incompatible con las predicciones de la MC, lo que representa un gran avance en el empeño por dilucidar el significado del experimento EPRB. Pero Bell fue más allá, pues, en principio, su desigualdad es verificable empíricamente. Este es el aspecto clave del trabajo de Bell: si la desigualdad es violada empíricamente, entonces la hipótesis del realismo local es falsa, y por tanto el entrelazamiento cuántico y las fantasmagóricas acciones a distancia son reales. A continuación, vamos a derivar una versión elemental de la desigualdad propuesta originalmente por Bell. En lo fundamental, la derivación sigue la misma lógica y se basa en los mismos supuestos que el razonamiento utilizado por Bell en su famoso artículo de 1964.

\section{Una derivación simple de la desigualdad de Bell}

Generalicemos el experimento EPRB imaginando que Alice y Bob pueden efectuar medidas conjuntas de spin sobre tres ejes arbitrarios $a$, $b$ y $c$ que no forman necesariamente ángulos rectos entre sí (Fig. 3). Aunque en la práctica resulta imposible medir el spin de cada protón sobre los ejes $a$, $b$ y $c$ simultáneamente, si se aceptan las reglas del experimento EPRB es posible asociar componentes de spin simultáneas a cada protón sobre los tres ejes, con independencia de la medición.\\

\begin{figure}[h]
  \centering
    \includegraphics[width=0.65\textwidth]{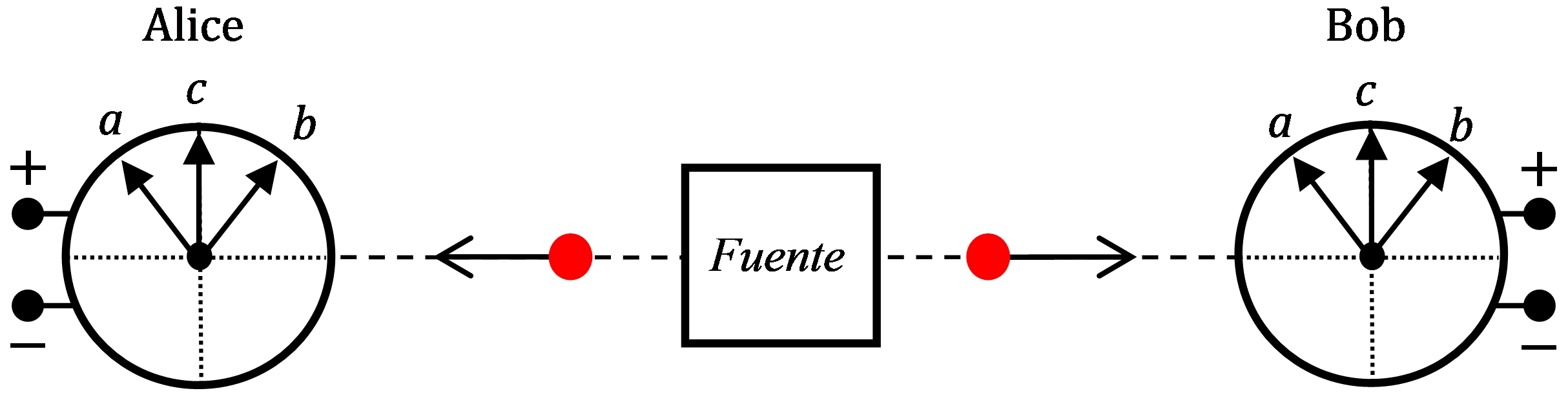}
  \caption{Dos protones en estado singlete (círculos rojos) viajan a largo de un mismo eje en sentidos opuestos hacia Alice y Bob, que pueden medir las componentes de spin sobre tres ejes arbitrarios, $a$, $b$ y $c$. Cada vez que Alice obtiene el resultado $+/–$, Bob obtiene el resultado $–/+$. }
\end{figure}

Siguiendo la lógica del experimento EPRB, supongamos que la fuente comienza a emitir secuencialmente un gran número de pares de protones en estado singlete. Mientras viajan hacia Alice y Bob, los protones no sufren ninguna perturbación. Como sabemos, las componentes de spin para cada eje o dirección solo pueden tomar los valores $+$ y $-$. Imaginemos que lo ejes de Alice y Bob nunca coinciden, pero cambian cada cierto número de mediciones de spin, de manera que por momentos Alice mide en la dirección $a$, y Bob mide en la dirección $b$, en otros momentos Alice mide en la dirección $c$ y Bob en la $b$, etc.\\

Dado que ahora tenemos tres ejes de medición, y para cada eje existen dos resultados, $+$ y $-$, se sigue que hay $2^{3} = 8$ resultados posibles, es decir, tenemos 4 resultados más que en el caso de las Tablas 1 y 2. Estos ocho resultados dan origen a los ocho conjuntos que aparecen en la Tabla 3. Al igual que en la Tabla 2, las medidas de componentes de spin revelan una correlación negativa perfecta. Cada conjunto tiene una cardinalidad, que establece la cantidad de pares de protones en estado singlete con las componentes de spin indicadas. El conjunto con cardinalidad $n_{1}$ reúne todos los pares, donde el protón detectado por Alice tiene componentes $a+,b+,c+$ y el protón detectado por Bob tiene componentes $a-,b-,c-$; el conjunto con cardinalidad $n_{2}$ reúne todos los pares, donde el protón detectado por Alice tiene componentes $a+,b+,c-$ y el protón detectado por Bob tiene componentes $a-,b-,c+$, etc. Si $N$ es la cardinalidad de los ocho conjuntos de la Tabla 3, que es equivalente al número total de pares de protones, debe cumplirse que:

\begin{equation}
N = n_{1} + n_{2} + n_{3}+...+n_{8} = \sum_{i=1}^{8} n_{i}. 
\end{equation}

Para derivar la desigualdad de Bell, necesitamos una notación que represente el número de medidas conjuntas de componentes de spin efectuadas por Alice y Bob. Como la desigualdad debe relacionar cantidades verificables empíricamente, también necesitamos que dicha notación de cuenta de componentes que puedan ser efectivamente medidas. Es decir, las componentes de spin inferidas no pueden formar parte de la ecuación. Como la MC prohíbe medir dos o más componentes de spin simultáneamente, y dado que los experimentos respaldan esta prohibición, la notación debe incorporar solo una componente de spin para cada protón de un par dado. Para cumplir con este requisito vamos a utilizar la notación $n(Alice,Bob)$, donde a la izquierda de la coma ponemos la componente de spin medida por Alice, y a la derecha la componente medida por Bob. Así, $n(a+,b+)$ representa el número de medidas conjuntas de spin donde Alice mide $a+$ y Bob mide $b+$.  A partir de la Tabla 3 observamos que los pares de protones con estas componentes de spin pertenecen al conjunto 3, cuya cardinalidad es $n_{3}$, y al conjunto 4, cuya cardinalidad es $n_{4}$, de manera que, 

\begin{equation}
n_{3}+n_{4}=n(a+,b+).
\end{equation}

Aplicando el mismo razonamiento y la misma notación, a partir de la Tabla 3 concluimos que:

\begin{equation}
n_{2}+n_{4} = n(a+,c+), 
\end{equation}

\begin{equation}
n_{3}+n_{7} = n(c+,b+).
\end{equation}

Por otra parte, podemos escribir la siguiente desigualdad, que no requiere demostración:

\begin{equation}
 n_{3}+n_{4} \leq \left( n_{2}+n_{4} \right) + \left( n_{3}+n_{7}\right)
\end{equation}

Si al lector no le parece evidente esta desigualdad, note que al tomar $n_{2}=n_{7}=0$ obtenemos una identidad. Reemplazando en esta desigualdad las Ecs. (3), (4) y (5) obtenemos,

\begin{equation}
n(a+,b+) \leq n(a+,c+)+n(c+,b+).
\end{equation}

Esta es la desigualdad WD, y es una de las formulaciones más simples del teorema de Bell. A partir de la Tabla 3, y siguiendo el mismo procedimiento descrito antes, es posible construir múltiples expresiones análogas a la Ec. (7) que relacionan otros pares de componentes de spin. Esta desigualdad fue deducida bajo la lógica del caso de Bertlmann, y por lo tanto satisface el realismo local. Vale decir, la Ec. (7) se basa en la hipótesis de que el realismo local es correcto, pues la hemos deducido poniendo en pie de igualdad las medidas e inferencias de Alice y Bob, tal como hicimos en el caso del experimento EPRB. Pese a ello, para un par de protones dado, cada término de la ecuación considera solo una componente de spin para cada protón, lo que significa que es verificable empíricamente. \\ 

\begin{table}[htbp]
\begin{center}
\caption{Componentes de spin en las direcciones $a$, $b$, $c$.}
\begin{tabular}{l l l} 
\toprule
Cardinalidad & Alice & Bob \\
\toprule
$n_{1}$ & $a+,b+,c+$ & $a-,b-,c-$ \\ 
\midrule
$n_{2}$ & $a+,b+,c-$ & $a-,b-,c+$ \\ 
\midrule
$n_{3}$ & $a+,b-,c+$ & $a-,b+,c-$ \\
\midrule
$n_{4}$ & $a+,b-,c-$ & $a-,b+,c+$\\
\midrule
$n_{5}$ & $a-,b+,c+$ & $a+,b-,c-$ \\
\midrule
$n_{6}$ & $a-,b+,c-$ & $a+,b-,c+$ \\
\midrule
$n_{7}$ & $a-,b-,c+$ & $a+,b+,c-$ \\
\midrule
$n_{8}$ & $a-,b-,c-$ & $a+,b+,c+$ \\
\bottomrule
\end{tabular}
\label{Componentes de spin en las direcciones z y x}
\end{center}
\end{table}

Dado que las predicciones cuánticas son estadísticas, para que la Ec. (7) pueda ser contrastada con la MC es necesario convertirla en una desigualdad entre probabilidades conjuntas; con este objetivo, debemos dividir ambos miembros por el número total $N$ de pares de protones, definido por la Ec. (2). Como $N > 0$, la desigualdad mantiene su sentido: 

\begin{equation}
\frac{n(a+,b+)}{N} \leq \frac{n(a+,c+)}{N} + \frac{n(c+,b+)}{N}.
\end{equation}

Si suponemos que $N$ es grande y que los pares de protones han sido emitidos de forma aleatoria desde la fuente, cada término de la Ec. (8) corresponde a la definición de probabilidad conjunta, vale decir, cada término expresa una proporción, definida como el cociente entre el número de eventos de interés y el número de eventos totales:

\begin{equation}
p(a+,b+) \leq p(a+,c+)+p(c+,b+).
\end{equation}

Esta ecuación nos dice que, si aceptamos la hipótesis del realismo local, la probabilidad de que Alice mida $a+$ y Bob mida $c+$, sumado a la probabilidad de que Alice mida $c+$ y Bob mida $b+$ siempre es mayor o igual que la probabilidad de que Alice mida $a+$ y Bob mida $b+$. Mediante algunos cálculos simples, en la siguiente sección contrastaremos la desigualdad de Bell con las predicciones de la MC.

\section{La MC y la violación de la desigualdad de Bell}

Con el propósito de contrastar la desigualdad WD con las predicciones estadísticas de la MC, conviene reescribir la Ec. (9) del siguiente modo:

\begin{equation}
 p(a+,c+)+p(c+,b+)- p(a+,b+) \geq 0.
\end{equation}

\begin{figure}[h]
  \centering
    \includegraphics[width=0.2\textwidth]{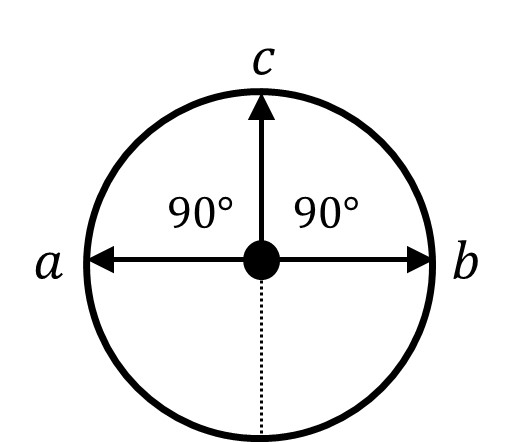}
  \caption{Orientaciones relativas de los ejes $a$, $b$ y $c$ para las medidas de spin bajo las reglas de la paradoja EPRB, donde se cumple que $\theta_{ac} = \theta_{cb} = 90^{o}$ y $\theta_{ab} = 180^{o}$.}
\end{figure}

A partir de las reglas de la MC se puede demostrar que las probabilidades conjuntas de la Ec. (10) vienen dadas por las siguientes expresiones [24]:

\begin{equation}
\begin{split}
p(a+,c+)=\frac{1}{2} sen^{2}\left( \frac{\theta_{ac}}{2} \right), \\ 
p(c+,b+)=\frac{1}{2} sen^{2}\left( \frac{\theta_{cb}}{2} \right), \\ 
p(a+,b+)=\frac{1}{2} sen^{2}\left( \frac{\theta_{ab}}{2} \right),
\end{split}
\end{equation}

donde $\theta_{ac}$ es el ángulo que forman los ejes $a$ y $c$, $\theta_{cb}$ es el ángulo que forman los ejes $c$ y $b$, y $\theta_{ab}$ es el ángulo que forman los ejes $a$ y $b$. A simple vista no es posible determinar si la MC predice resultados experimentales distintos de los que predice el experimento EPRB. Para verificar esto último, es necesario reemplazar las Ecs. (11) en la Ec. (10):

\begin{equation}
sen^{2}\left( \frac{\theta_{ac}}{2} \right)+ sen^{2}\left( \frac{\theta_{cb}}{2} \right)- sen^{2}\left( \frac{\theta_{ab}}{2} \right) \geq 0.
\end{equation}

\begin{figure}[h]
  \centering
    \includegraphics[width=0.2\textwidth]{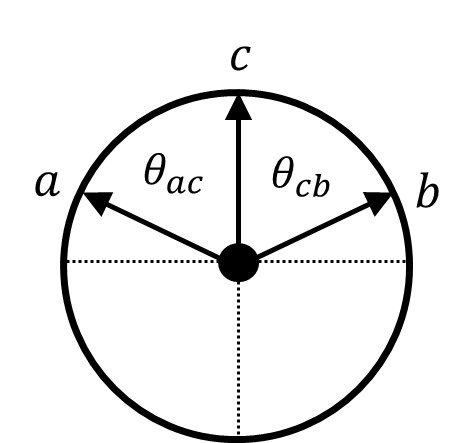}
  \caption{Orientaciones de los ejes $a$, $b$ y $c$ para las medidas de spin bajo las condiciones más generales impuestas por la desigualdad de Bell, donde los ángulos relativos son distintos de $0^{o}$ y $90^{o}$.}
\end{figure}

Como es fácil comprobar, si se escogen los ángulos de modo que $\theta_{ac} = \theta_{cb} = 90^{o}$ y $\theta_{ab}=180^{o}$, no se genera conflicto con la Ec. (12) (ver Fig. 4). En efecto, al introducir estos valores en la Ec. (12) se obtiene

\begin{equation}
 sen^{2}\left( 45^{o} \right)+ sen^{2}\left( 45^{o} \right)- sen^{2}\left( 90^{o} \right) \geq 0.
\end{equation}

Evaluando cada término en esta ecuación resulta $0 \geq 0$, lo que siempre es verdadero. Vemos entonces que si los ejes $a$, $b$, $c$ están alienados o forman ángulos rectos entre sí, la desigualdad de Bell no resulta violada, aun cuando los protones estén entrelazados. Aunque este punto fue señalado en la sección anterior vale la pena enfatizarlo: el escenario del experimento EPRB, que es equivalente al caso de Bertlmann, es indistinguible del escenario de entrelazamiento cuántico cuando los ejes de medición son colineales o perpendiculares. Pero Bell descubrió que cuando ello no sucede, estos dos escenarios conducen a predicciones distintas. \\ 

\begin{figure}[h]
  \centering
    \includegraphics[width=0.7\textwidth]{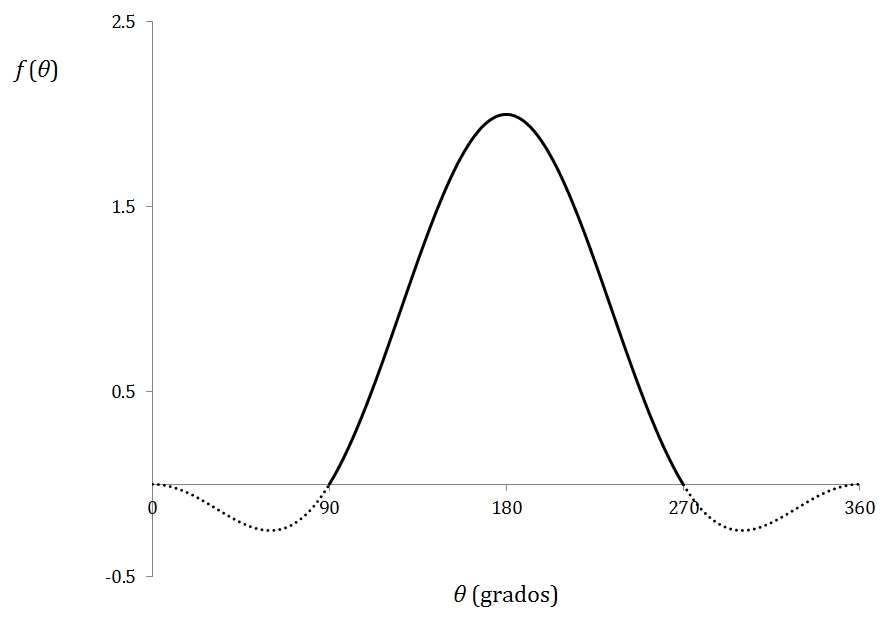}
  \caption{La desigualdad de Bell es violada cuando la curva se encuentra bajo el eje $f(\theta) = 0$, lo que sucede para todo ángulo $\theta$ tal que $0 < \theta < 90^{o}$ o $270^{o} < \theta < 360^{o}$.}
\end{figure}

Siguiendo a Bell, consideremos la situación ilustrada en la Fig. 5, donde los ángulos relativos entre los ejes $a$, $b$ y $c$ son distintos de $0^{o}$ y $90^{o}$. Escojamos los ángulos de modo que $\theta_{ac} = \theta_{cb} = \theta$ y $\theta_{ab}=2\theta$. En estas condiciones, la Ec. (12) puede escribirse como:

\begin{equation}
2sen^{2}\left( \frac{\theta}{2} \right) - sen^{2}(\theta) \geq 0.
\end{equation}

Si en la Ec. (14) tomamos $\theta = 45^{o}$ se obtiene $0 \geq 0,207$ de modo que la desigualdad es violada. Se puede demostrar que la Ec. (14) será violada para todo ángulo tal que $0 < \theta < 90^{o}$ o $270^{o} < \theta < 360^{o}$. Para verificar este hecho gráficamente, a partir de la Ec. (14) conviene definir la siguiente función:

\begin{equation}
f(\theta) = 2sen^{2}\left( \frac{\theta}{2} \right) - sen^{2}(\theta).
\end{equation}

Naturalmente, cualquier valor de $f(\theta)$ menor que cero implica una violación de la desigualdad WD. Al graficar la Ec. (15) poniendo $f(\theta)$ en el eje de las ordenadas, y $\theta$ (en grados) en el eje de las abscisas, se observa claramente que la curva se encuentra bajo el eje $f(\theta)$ para todos los ángulos tales que $0 < \theta < 90^{o}$ o $270^{o} < \theta < 360^{o}$ (ver Fig. 6). Las implicancias de este resultado son claras: la MC no satisface el realismo local. En la siguiente sección discutiremos con mayor detalle el significado físico de este resultado, para lo cual analizaremos la desigualdad de Bell en el contexto de la evidencia empírica.

\section{Las fantasmagóricas acciones a distancia y la evidencia experimental}

Para entender el significado y los alcances de los experimentos que han permitido establecer la violación de la desigualdad de Bell, consideremos el primer experimento de este tipo y el más simple conceptualmente, realizado en 1972 por John Clauser, uno de los galardonados con el premio nobel de física 2022 por su trabajo pionero en este campo. Tanto en este experimento como en los que le sucedieron, no se contrastó la versión original de la desigualdad de Bell ni tampoco la desigualdad WD debido a que estos resultados requieren condiciones experimentales idealizadas que son muy difíciles de alcanzar en un laboratorio. Clauser utilizó una generalización del teorema de Bell conocida como \textit{desigualdad CHSH}, que derivó en 1969 junto con Michael Horne, Abner Shimony y Richard Holt [32]. Esta desigualdad facilita la contrastación experimental al permitir el uso de dispositivos de medición imperfectos. \\

En su experimento, que realizó en colaboración con Stuart Freedman, Clauser utilizó pares de fotones entrelazados a los que midió su polarización lineal, una propiedad análoga al spin de los pares de protones entrelazados que analizamos antes [8]. Clauser y Freedman instalaron una fuente de fotones en medio de dos detectores destinados a determinar la tasa de coincidencias, que consiste en medir dos fotones que golpean los detectores simultáneamente. Como fuente de partículas, el experimento utilizó átomos de calcio que podían emitir fotones entrelazados después de haberlos iluminado con una luz especial. Como se lustra en la Fig. 7, cada detector estaba equipado con un polarizador que se podía girar en una orientación arbitraria. La función de los sistemas de detectores y polarizadores es equivalente a las de los dispositivos para medir spin que ya conocemos. El efecto de los polarizadores sobre los fotones es análogo al que provocan los lentes de Sol, que bloquean la luz que ha sido polarizada en un cierto plano, por ejemplo, al reflejarse en el agua. Para resaltar las analogías entre las medidas de spin sobre pares de protones y las medidas de polarización sobre pares de fotones, al sistema del polarizador y detector ubicado a la derecha le llamaremos Alice ($A$), y al sistema del polarizador y detector ubicado a la izquierda le llamaremos Bob ($B$). El estado de polarización de los fotones entrelazados con respecto a un eje fijo se escribe como:

\begin{equation}
\vert \psi \rangle =\frac{1}{\sqrt{2}} \vert + \rangle_{A} \vert + \rangle_{B} + \frac{1}{\sqrt{2}}\vert - \rangle_{A} \vert - \rangle_{B}.
\end{equation}

donde $+$ representa la polarización vertical, y $-$ representa la polarización horizontal. La polarización lineal es una propiedad dicotómica, al igual que el spin. Notemos la semejanza entre esta igualdad y la Ec. (1). La única diferencia es que esta última ecuación predice una correlación positiva perfecta entre medidas de polarización sobre un mismo eje, mientras que la Ec. (1) predice una correlación negativa perfecta. La Ec. (16) indica que los fotones están en un estado entrelazado compuesto por dos alternativas excluyentes que describen las probabilidades de una medida de polarización efectuada por Alice y Bob en una dirección fija arbitraria. El caso en que la polarización de ambos fotones es vertical se simboliza $\vert +_{A} \rangle \vert + \rangle_{B}$, y el caso en que la polarización de ambos fotones es horizontal se simboliza $\vert - \rangle_{A} \vert - \rangle_{B}$; según la regla de Born, la probabilidad de que ocurra cualquiera de estas opciones es $1/2$.\\  

Imaginemos que en el experimento se envía un par de fotones hacia Alice y Bob, cuyos polarizadores están orientados en la misma dirección, que supondremos corresponde a la vertical. Como revela la Ec. (16), si uno de los fotones atraviesa el polarizador de Alice, entonces el otro atravesará el polarizador de Bob. Si los polarizadores están en ángulo recto entre sí, un fotón será bloqueado mientras que el otro podrá atravesar. Al igual que en los experimentos con protones analizados en la sección anterior, el truco del experimento de Clauser consiste en efectuar las medidas con los polarizadores desplazados entre sí en ángulos distintos de $0^{o}$ y $90^{o}$. Bajo estas condiciones, los resultados pueden variar: a veces ambos fotones atraviesan los polarizadores, a veces solo uno y, a veces, ninguno. La frecuencia con la que ambos fotones atraviesan depende del ángulo entre ellos. \\  

\begin{figure}[h]
  \centering
    \includegraphics[width=0.7\textwidth]{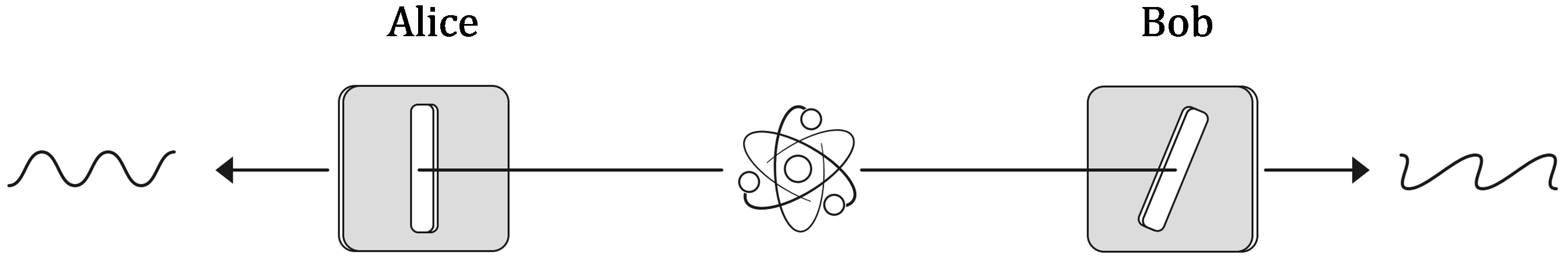}
  \caption{El experimento de Clauser usó átomos de calcio que podían emitir fotones entrelazados después de haberlos iluminado con una luz especial. Los resultados Bob demostraron que se viola la desigualdad de Bell (Fuente: https://www.nobelprize.org/prizes/physics/2022/popular-information/).}
\end{figure}

A partir de la desigualdad CHSH, Clauser demostró teóricamente que la máxima violación se produce cuando los polarizadores están desplazados entre sí en $22,5^{o}$ o $67,5^{o}$ [31]. Los investigadores recopilaron 200 horas de datos y encontraron que las tasas de coincidencia violaron la desigualdad de Bell. Sin embargo, pese a su innegable importancia, este resultado dista mucho de ser concluyente, ya que el experimento de Clauser y Freedman presenta tres grandes debilidades, llamadas habitualmente \textit{lagunas} (loopholes), que permiten objetar la conclusión de que se ha producido una violación de la hipótesis del realismo local. Estas lagunas imitan el entrelazamiento cuántico, aun cuando los pares de fotones sean entidades físicas independientes [33]. Veamos brevemente en qué consisten estas lagunas.\\  

La primera, denominada \textit{laguna de localidad} (locality loophole) surge del hecho que, en virtud de la localidad, la desigualdad de Bell asume que las medidas de Alice no influyen sobre las medidas de Bob, y viceversa [34]. Sin embargo, dado que Clauser y Freedman utilizaron polarizadores que permanecían en un ángulo fijo durante largos períodos, es posible invocar la presencia de alguna interacción entre los polarizadores y la fuente, que se desplaza a una velocidad igual o menor que la de la luz. Esta interacción podría imitar los efectos no locales del entrelazamiento cuántico, aun cuando se trata de interacciones locales, que se propagan a velocidades inferiores a la de la luz. Luego, un modelo realista local que invoque la presencia de este tipo de interaccione locales, podría reproducir los resultados del experimento. Para evitar la laguna de la localidad es necesario que la elección de la orientación de los polarizadores se haga mientras los fotones están en vuelo desde la fuente, pues con ello se excluye la posibilidad de que existan interacciones superluminales entre los dispositivos que imiten el entrelazamiento cuántico. En la jerga relativista, esto significa que el intervalo entre los eventos de elección de la orientación de los polarizadores debe ser de tipo espacio.\\  

La segunda laguna, conocida como \textit{laguna de detección} (detection loophole) proviene del hecho que los dispositivos de detección de fotones son imperfectos, y por lo tanto, ninguno tiene una eficiencia de un 100\% [34]. Si se detectan muy pocos fotones, la medida podría estar recogiendo una muestra no representativa que sesga las correlaciones, y que imita el entrelazamiento cuántico. Por consiguiente, un modelo realista local que invoque esta laguna podría reproducir los resultados del experimento. Como es evidente, para evitar esta laguna es necesario disponer de detectores de fotones más eficientes, que recojan una muestra representativa de pares de fotones correlacionados. De hecho, se puede demostrar que la detección de más de las tres cuartas partes de los fotones soluciona esta laguna.\\  

La tercera laguna es por lejos la más compleja, y surge ante la posibilidad de que el libre albedrío sea una ilusión [34]. Este es el denominado \textit{superdeterminismo}, un concepto que supone que todo en el universo está predeterminado. Además de basarse en el realismo local, el teorema de Bell supone que no habitamos un universo superdeterminista, lo que significa que las medidas realizadas en cada detector pueden elegirse independientemente unas de otras. Esta libertad de elección se denomina \textit{independencia de medición} o \textit{independencia estadística}. El superdeterminismo implica que Clauser y Freedman no tuvieron libertad para escoger su configuración experimental, lo que se traduce en una violación del supuesto de independencia estadística. Llevado al extremo, el superdeterminismo sugiere que cada evento en todo el espacio-tiempo está determinado por las condiciones iniciales en el Big Bang. En un universo superdeterminista, las correlaciones del experimento de Clauser y Freedman podrían ocurrir aun cuando no exista entrelazamiento, y un modelo realista local que invoque el superdeterminismo sería capaz de imitar los resultados esperables en un escenario no local. En estricto rigor, es imposible evadir la laguna del superdeterminismo, pero se pueden suavizar sus efectos si las medidas realizadas son elegidas mediante generadores de números aleatorios [35].\\  

En la década de 1980, Bell habló sobre el superdeterminismo en una entrevista con la BBC que le realizó el físico Pauli Davies, y que luego fue transcrita y editada [36]. En la entrevista, Bell señala que es posible evitar los efectos no locales y las fantasmagóricas acciones a distancia si suponemos que en el universo impera un determinismo absoluto o superdeterminismo. Según Bell, suponer esto implica que no solo la naturaleza inanimada está regida por un mecanismo de relojería, sino que incluso nuestro comportamiento estaría completamente determinado, y por lo tanto la creencia de que tenemos libertad para decidir dónde y cómo realizar experimentos es una ilusión. En un universo superdeterminista no hay necesidad de señales superluminales que le comuniquen a la partícula de Alice que la partícula de Bob fue medida, porque el universo, incluida la partícula de Alice, “conoce” esa medida y su resultado. Bell también argumentaba que la laguna del superdeterminismo es inverosímil, y que las medidas elegidas mediante generadores de números aleatorios permiten asumir que las elecciones de los experimentalistas son libres [36]. Zeilinger tampoco considera verisímil el superdeterminismo, señalando que, si fuera cierto, algunas de sus implicaciones pondrían en tela de juicio el valor de la ciencia misma al destruir la falsabilidad\footnote{La falsabilidad, también llamada refutabilidad, es la capacidad de una teoría o hipótesis de ser sometida a potenciales pruebas que la refuten o falseen. Según el filósofo Karl Popper, la falsabilidad es el principal criterio que permite distinguir la ciencia de la pseudociencia.} [37]. Como cabe esperar, no todos los especialistas están de acuerdo con los puntos de vista de Bell y Zeilinger. En cualquier caso, el debate continúa, y por de pronto se ve difícil que se alcance algún consenso.\\   

Estas tres lagunas en el experimento de Clauser y Freedman impulsaron a los otros dos galardonados con el nobel de física 2022, Aspect y Zeilinger, a realizar experimentos libres de lagunas (loophole-free). Aunque los experimentos de Aspect y Zeilinger no lograron alcanzar plenamente su cometido, allanaron el camino para que otros experimentos pudiesen llegar a la meta. El camino al éxito fue largo y difícil. Para tener una idea de las dificultades que enfrentaron quienes siguieron el camino trazado por Clauser, Aspect y Zeilinger, consideremos lo siguiente: la estrategia más simple y directa para evadir la laguna de la localidad es situar a Alice y Bob a una gran distancia mutua, de tal manera que la elección de la orientación de los polarizadores pueda hacerse mientras lo fotones están en vuelo. No obstante, al aumentar la distancia, se reduce la eficiencia de los detectores de fotones, lo que se traduce en un menor número de fotones registrados, facilitando que la laguna de detección entre en escena.\\   

Debieron transcurrir más de cuatro décadas desde el experimento pionero de Clauser y Freedman para que la tecnología estuviese lo suficientemente madura para permitir el diseño e implementación de un experimento libre de lagunas (salvo quizá el superdeterminismo). Esta hazaña fue realizada el año 2014 por un equipo liderado por Ronald Hensen de la Delft University of Technology. Aunque los detalles del experimento son complejos y no podemos discutirlos aquí, la idea básica está en excelente sintonía con el experimento EPRB y con la propuesta original de Bell [38]. En lugar de fotones entrelazados, Hensen y sus colaboradores emplearon pares de electrones entrelazados cuyas componentes de spin fueron medidas cuando las partículas se encontraban separadas una distancia de $1,3km$ [39].\\   

El aspecto central del experimento de Hensen es que se encuentra libre de lagunas, y por tanto permite contrastar directamente la hipótesis del realismo local. El resultado del experimento provee sólida evidencia en favor de los hallazgos de los experimentos que le precedieron: el realismo local es falso. \\

La no localidad se encuentra en flagrante conflicto con la versión fuerte del principio de relatividad de Einstein, según el cual ninguna señal o interacción puede propagarse más rápido que la luz en el vacío. Sin embargo, la no localidad es compatible con una versión débil de dicho principio, que solo prohíbe el envío de información macroscópica (utilizable por personas) con rapidez superior a la de la luz. En efecto, se ha demostrado de forma fehaciente que las correlaciones entre pares de partículas entrelazadas no permiten el envío de información. Sin embargo, como ha señalado el propio Bell, es evidente que el entrelazamiento cuántico conlleva un conflicto con la física relativista [35]. Probablemente este conflicto fue la principal razón que llevó a Einstein a rechazar el entrelazamiento, y a calificarlo como una fantasmagórica acción a distancia. En cualquier caso, la evidencia empírica deja poco espacio para la duda: la acción a distancia forma parte de la realidad física, y los constituyentes microscópicos del universo no siempre pueden considerarse como entidades físicas independientes.

\section{Realismo versus positivismo}
En la introducción hemos señalado que el realismo no es la única postura filosófica posible frente a la desigualdad de Bell y a los resultados experimentales. Para entender esta idea, recordemos que el realismo local contiene en verdad dos hipótesis: el realismo y la localidad. Cabe preguntarse, por tanto, a cuál de estas hipótesis nos obliga a renunciar la evidencia experimental. No es una pregunta que admita una respuesta fácil. Más atrás hemos sugerido que los experimentos implican que la localidad es falsa, pero debemos ser cuidadosos, pues la veracidad de esta premisa es menos evidente de lo que podría parecer en primera instancia. Desde un punto de vista estrictamente lógico tenemos tres opciones: o bien el realismo es falso, o bien la localidad es falsa, o bien ambas opciones son falsas. \\

John S. Bell enfrentó estas opciones con pragmatismo: Renunciar al realismo es una postura demasiado costosa y radical, en especial, considerando que para explicar la violación de la desigualdad de Bell nos basta con abandonar la localidad [35]. Si optamos por conservar el realismo y renunciamos a la localidad, podemos afirmar que el experimento de Hensen y sus colaboradores es consistente con la idea de que las partículas cuánticas se comunican instantáneamente mediante interacciones superluminales, sin importar cuan alejadas se encuentren una de otra. Este fenómeno es el entrelazamiento cuántico, y de todas las peculiaridades de la MC quizá sea la más radical y la que revela de forma más evidente la ruptura definitiva entre la física clásica y el mundo cuántico.\\

Sin embargo, es perfectamente válido no querer entrar en los debates filosóficos que conlleva la hipótesis del realismo local. Asumir esta postura implica, en la práctica, parapetarse en el formalismo matemático de la MC y en los resultados experimentales, y no hacer ninguna afirmación que vaya más allá de este reducido escenario. Existen diversas posturas filosóficas que, en general, estarían de acuerdo con este planteamiento. Desde el punto de vista del desarrollo histórico de la MC, la  mas importante es el \textit{positivismo}. En términos amplios, el positivismo sostiene que todo conocimiento genuino se limita a la interpretación de los hallazgos “positivos”, es decir, empíricamente verificables. En el contexto de la MC, el positivismo afirma que no estamos autorizados a hacer afirmaciones que vallan más allá de los resultados experimentales y del formalismo cuántico [36]. \\

Situado en el problema que nos ocupa, el positivismo sostiene que lo único indiscutible es que la violación experimental de la desigualdad de Bell revela la existencia de correlaciones estadísticas entre pares de partículas en estado singlete, que es un tipo de \textit{estado no factorizable}. El realismo va más allá, y asegura que las correlaciones ponen de manifiesto el entrelazamiento o conexión física que existe entre los pares de partículas, lo que se traduce matemáticamente en el hecho que el estado singlete es no factorizable. En el contexto del experimento EPRB, cuando hablamos de un estado no factorizable, queremos decir que la función de onda del par de partículas no puede separarse o factorizarse en una función $\vert \psi \rangle_{A}$ para el protón que viaja hacia Alice, y otra función $\vert \psi \rangle_{B}$ para el protón que viaja hacia Bob. Si $\vert \psi \rangle$ es la función de onda para el estado singlete, esta idea se expresa formalmente como [45]:

\begin{equation}
\vert \psi \rangle \neq \vert \psi \rangle_{A} \otimes \vert \psi \rangle_{B}.
\end{equation}

Los positivistas se refieren al estado singlete como un <<estado no factorizable>>, en tanto que los realistas prefieren el término <<estado entrelazado>>. Vemos entonces que el positivista es cauto, y opta por utilizar una expresión cuyo significado está circunscrito al terreno formal y experimental, mientras que el realista es más audaz, y utiliza un término que alude a una conexión física entre pares de partículas, que no es observable en forma directa. Para el realista, la imposibilidad matemática de factorizar la función de onda tiene como correlato empírico la imposibilidad de separar físicamente a las partículas, que se comportan como una entidad única e indivisible, con independencia de la distancia que las separe.\\

En síntesis, dependiendo de la posición filosófica que tomemos, podemos interpretar la violación experimental de la desigualdad de Bell de manera diferente o, si se prefiere, con distinto alcance. No obstante, en los planos formal y empírico las posturas positivista y realista son coincidentes, pues nadie puede negar la existencia de efectos no locales, que matemáticamente se expresan en la imposibilidad de factorizar la función de onda del estado singlete, y que empíricamente se ponen de manifestó en las correlaciones estadísticas observadas.\\

El  más importante exponente del positivismo como postura filosófica frente a los problemas interpretativos de la MC fue el físico teórico danés Niels Bohr [36]. La posición de Bohr ha llegado a conocerse como \textit{interpretación de Copenhague}, ya que fue desarrollada principalmente en Copenhague, la capital de Dinamarca. Einstein, por su parte, era realista, lo que lo llevó a sostener un largo debate con Bohr sobre los fundamentos de la MC. La interpretación de Copenhague es bastante sutil, y un análisis detallado de esta posición excede los alcances de este trabajo. En todo caso, desde un punto de vista operacional, la interpretación de Bohr nos dice que solo podemos conocer la parte de la naturaleza que nuestros experimentos ponen de manifiesto. Como cabe esperar de un realista, Einstein nunca aceptó la interpretación de Copenhague. Aunque Einstein y Bohr abandonaron este mundo hace mucho tiempo, el debate que abrieron continua vivo, y el enfrentamiento entre realistas y positivistas en torno al significado de la MC no da señales de tregua.\\

Entre los físicos que trabajan activamente en problemas relacionados con el mundo cuántico, la postura mayoritaria u ortodoxa parece ser el positivismo, porque permite evitar la controversia sobre el significado de la MC y de la violación de la desigualdad de Bell. No cabe duda de que se trata de una postura cómoda, que soslaya por decreto las dificultades interpretativas de la MC, pero debemos reconocer que en algunos aspectos, esta postura ha resultado provechosa, al evitar que los investigadores se enreden en problemas filosóficos que podrían distraerlos de sus labores estrictamente científicas. Pero también debemos reconocer que, desde un punto de vista educativo, esta postura es profundamente insatisfactoria, pues en vez de promover el debate y el libre intercambio de ideas, impide, por decreto, que los estudiantes planteen preguntas incomodas que muchos profesores no están preparados para responder. Esta falta de preparación no se debe sólo a las dificultades inherentes al tema, sino al hecho mas básico de que una proporción significativa de quienes enseñan física a nivel universitario no son conscientes de su posición filosófica frente a la interpretación de la MC.\\

Pero no podemos ser demasiado críticos con los profesores por esta falta de conciencia, pues hay que reconocer que los problemas interpretativos de la MC son profundos y complejos, y exceden largamente la formación típica de un físico. Además, no todos disponemos del tiempo, la preparación y el ánimo de sumergirnos en estos problemas. Para dar una idea de la complejidad del tema basta señalar que existen múltiples interpretaciones de la MC, donde el realismo y el positivismo son solo la punta del iceberg. En cualquier caso, independientemente de los problemas de interpretación, la MC ha podido explicar esencialmente todo el mundo microscópico y ha hecho predicciones que han sido probadas experimentalmente con un alto grado de precisión y exactitud, por lo que es una teoría unánimemente aceptada.\\

En este escenario, es de esperar que este trabajo contribuya a generar el tan necesario debate en las aulas universitarias y en los cursos de física en torno a la desigualdad de Bell, la no localidad, el entrelazamiento cuántico, la interpretación de Copenhague, así como en torno a otros fascinantes y controvertidos temas relacionados con los fundamentos de la MC.

\section{Comentarios finales}

Han trascurrido más de tres décadas desde la temprana muerte de John S. Bell, ocurrida en 1990 ha causa de una hemorragia cerebral, pero no cabe duda de que estaría más que satisfecho con los extraordinarios avances producidos en el campo que el inauguró. La temprana muerte de Bell impidió que recibiera el premio nobel de física, al que fue nominado el mismo año de su muerte [40]. Por lo tanto, el galardón entregado a Clauser, Aspect y Zeilinger es también un reconocimiento al trabajo pionero de Bell y a su notable teorema, que ha revolucionado nuestra comprensión del universo cuántico.\\ 

\begin{figure}[h]
  \centering
    \includegraphics[width=0.5\textwidth]{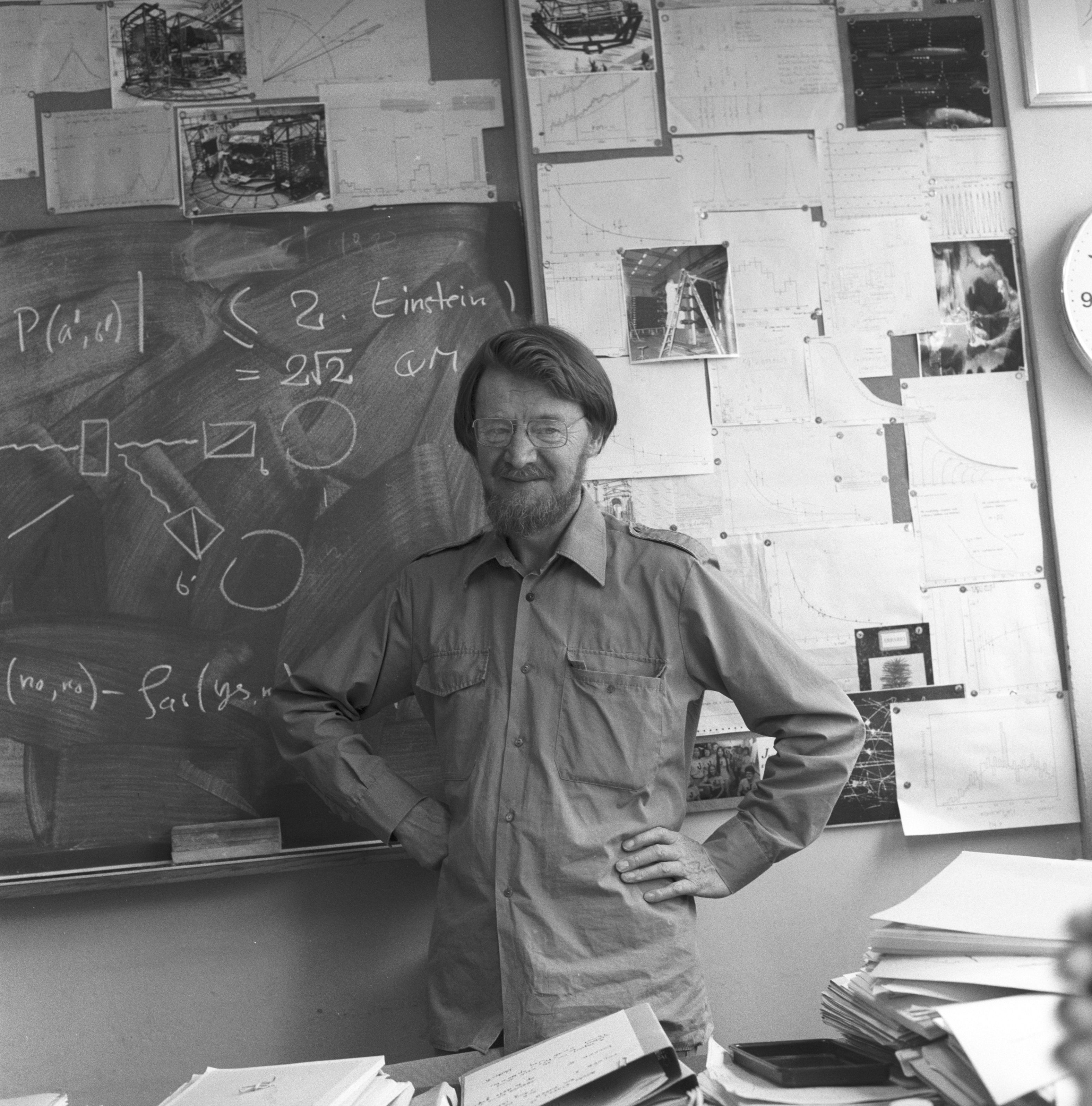}
  \caption{Figura 8. John S. Bell en su oficina en CERN, junio de 1982 (Fuente: https://cds.cern.ch/record/1823937).}
\end{figure}

El teorema de Bell ha sido calificado como el descubrimiento más profundo de la ciencia [41]. Aunque las motivaciones originales de Bell eran principalmente filosóficas, y probablemente nunca imaginó que sus esfuerzos por comprender los fundamentos de la mecánica cuántica tendrían tanto impacto, lo cierto es que su trabajo ha abierto nuevos horizontes científicos y tecnológicos que recién comenzamos a explorar (ver Fig. 8). Entre las aplicaciones más vanguardistas del trabajo Bell se encuentran la criptografía cuántica, la computación y la teletransportación cuánticas [42]. Para honrar el legado científico de Bell, en el año 2009 la Universidad de Toronto estableció el Bell Prize (\textit{The John Stewart Bell Prize for Research on Fundamental Issues in Quantum Mechanics and their Applications}) que está destinado a reconocer los principales avances relacionados con los fundamentos de la mecánica cuántica.\\

Los resultados de esta auténtica revolución también han llegado a los laboratorios de pregrado [43–49], ya que durante las décadas transcurridas desde los trabajos pioneros de Clauser y sus colaboradores, los experimentos con partículas correlacionadas se han hecho cada vez más simples y accesibles, y hoy en día es habitual que los estudiantes de física de países desarrollados realicen estos experimentos rutinariamente como parte de su proceso formativo. Estos desarrollos han tenido un positivo impacto en la formación de los estudiantes y han conducido a un replanteamiento de la forma en que se enseña MC [49].\\

Pero no debemos olvidar que el inicio de esta revolución científica comenzó con el experimento EPR, que nació del inconformismo de Einstein ante la visión ortodoxa de la MC, y ante su deseo de entender en profundidad el mundo cuántico. Aunque en un comienzo el experimento EPR parecía destinado a convertirse en una reliquia científica sin ninguna conexión con el mundo real, terminó por transformarse en el fundamento del trabajo de Bell, y hoy en día el artículo de Einstein, Podolski y Rosen es uno de los más citados en la historia de la física.\\ 

Esto reivindica la figura de Einstein, que erróneamente muchos han visto como un reaccionario superado por la física de su tiempo. El artículo EPR demuestra lo contrario: Asumiendo que existe una realidad objetiva que es independiente del observador, Einstein y sus colaboradores descubrieron que en el corazón de la MC se ocultaba algo sorprendente que otros no habían visto: las fantasmagóricas acciones a distancia. El hecho que los experimentos hayan demostrado que dichas acciones forman parte del mundo microscópico no debe interpretarse como una derrota de Einstein sino como un triunfo del espíritu científico, que nunca deja de cuestionar, y que erige a la naturaleza en el único juez definitivo de los logros de la empresa científica. Al igual que Einstein y sus colaboradores, Bell no se conformó con la visión ortodoxa de la MC y decidió ir más allá. Solo el tiempo nos dirá que tan lejos nos conducirá el inconformismo de Einstein y Bell. 

\section*{Referencias}

[1] A. Einstein, M. Born, E. Born. The Born-Einstein Letters (Macmillan, London, 1971).

\vspace{2mm}

[2]	A. Einstein, B. Podolsky, N. Rosen, Can quantum-mechanical description of physical reality be considered complete?, Phys. Rev. \textbf{47}, 777 (1935) .

\vspace{2mm}

[3]	J.S. Bell, On the Einstein Podolsky Rosen Paradox, Physics \textbf{1}, 195 (1964).

\vspace{2mm}

[4]	W. Tittel, J. Brendel, H. Zbinden, N. Gisin, Violation of Bell inequalities by photons more than 10 km apart, Phys. Rev. Lett. \textbf{81}, 3563 (1998).

\vspace{2mm}

[5]	G. Weihs, T. Jennewein, C. Simon, H. Weinfurter, A. Zeilinger, Violation of Bell’s inequality under strict Einstein locality conditions, Phys. Rev. Lett. \textbf{81}, 5039 (1998).

\vspace{2mm}

[6]	D. Salart, A. Baas, J.A.W. van Houwelingen, N. Gisin, H. Zbinden, Spacelike Separation in a Bell Test Assuming Gravitationally Induced Collapses, Phys. Rev. Lett. \textbf{100}, 220404 (2008).

\vspace{2mm}

[7]	D. Bouwmeester, J.-W. Pan, K. Mattle, M. Eibl, H. Weinfurter, A. Zeilinger, Experimental quantum teleportation, Nature \textbf{390}, 575 (1997). 

\vspace{2mm}

[8]	S.J. Freedman, J.F. Clauser, Experimental Test of Local Hidden-Variable Theories, Phys. Rev. Lett. \textbf{28}, 938 (1972).

\vspace{2mm}
 
[9]	A. Aspect, J. Dalibard, G. Roger, Experimental Test of Bell`s Inequalities Using Time-Varying Analyzers, Phys. Rev. Lett. \textbf{49}, 1804 (1982).

\vspace{2mm}

[10]	J.S. Bell, Bertlmann’s socks and the nature of reality, Le Journal de Physique Colloques \textbf{42}, 41 (1981).

\vspace{2mm}

[11] J. Pinochet, D. Rojas-Líbano, Una demostración simple de la desigualdad de Bell basada en la teoría elemental de conjuntos, Rev. Bras. Ensino Fís. \textbf{38}, e3303-2 (2016). 

\vspace{2mm}

[12] E.P. Wigner, On hidden variables and quantum mechanical probabilities, Am. J. Phys. \textbf{38}, 1005 (1970).

\vspace{2mm}

[13] G. Blaylock, The EPR paradox, Bell’s inequality, and the question of locality, Am. J. Phys. \textbf{78}, 111 (2010).

\vspace{2mm}
 
[14] T. Maudlin, What Bell proved: A reply to Blaylock, Am. J. Phys. \textbf{78}, 121 (2010).

\vspace{2mm}
 
[15] R.B. Griffiths, EPR, Bell, and quantum locality, Am. J. Phys. \textbf{79}, 954 (2011). 

\vspace{2mm}

[16] M. Nauenberg, QBism AND LOCALITY IN QUANTUM MECHANICS, Am. J. Phys. \textbf{83}, 197 (2015).

\vspace{2mm}
 
[17] J. Bricmont, W\textit{hat Did Bell Really Prove}?, in: M. Bell, S. Gao (Eds.), Quantum Nonlocality and Reality: 50 Years of Bell’s Theorem (Cambridge University Press, Cambridge, 2016). 

\vspace{2mm}

[18] M.G. Alford, Ghostly action at a distance: A non-technical explanation of the Bell inequality, Am. J. Phys. \textbf{84}, 448 (2016).

\vspace{2mm}
 
[19] D.M. Greenberger, M.A. Horne, A. Shimony, A. Zeilinger, Bell’s theorem without inequalities, Am. J. Phys. \textbf{58}, 1131 (1990).

\vspace{2mm}
 
[20] F. Kuttner, B. Rosenblum, Bell’s Theorem and Einstein’s ‘Spooky Actions’ from a Simple Thought Experiment, Phys. Teach. \textbf{48}, 124 (2010).

\vspace{2mm}

[21] P.G. Kwiat, L. Hardy, The mystery of the quantum cakes, Am. J. Phys. \textbf{68}, 33 (2000).

\vspace{2mm}

[22] N.D. Mermin, Quantum mysteries revisited, Am. J. Phys. \textbf{58}, 731 (1990).

\vspace{2mm}

[23] K. Jacobs, H.M. Wiseman, An Entangled Web of Crime: Bell’s Theorem as a Short Story, Am. J. Phys. \textbf{73}, 932 (2005).

\vspace{2mm}

[24] L. Maccone, A simple proof of Bell’s inequality, Am. J. Phys. \textbf{81}, 854 (2013).

\vspace{2mm}

[25] J.J. Sakurai, \textit{Modern Quantum Mechanics} (Addison Wesley, USA, 1994).

\vspace{2mm}

[26] B. d’Espagnat, The quantum theory and reality, Scientific American \textbf{241}, 158 (1979).

\vspace{2mm}

[27] J. Marckwordt, A. Muller, D. Harlow, D. Franklin, R.H. Landsberg, Entanglement Ball: Using Dodgeball to Introduce Quantum Entanglement, Phys. Teach. \textbf{59}, 613 (2021). 

\vspace{2mm}

[28] D. Bohm, \textit{Quantum Theory} (Dover Publications Inc, New Jersey, 1989).

\vspace{2mm}

[29] C. Cohen-Tannoudji, B. Diu, F. Laloë, \textit{Quantum Mechanics} (Wiley, New York, 2019) V. 2.

\vspace{2mm}

[30] C. Cohen-Tannoudji, B. Diu, F. Laloe, \textit{Quantum Mechanics} (Wiley, New York, 2019) V. 1.

\vspace{2mm}

[31] N.D. Mermin, Is the Moon There When Nobody Looks? Reality and the Quantum Theory, Physics Today \textbf{38}, 38 (1985).

\vspace{2mm}

[32] J.F. Clauser, M.A. Horne, A. Shimony, R.A. Holt, Proposed Experiment to Test Local Hidden-Variable Theories, Phys. Rev. Lett. \textbf{23}, 880 (1969). 

\vspace{2mm}

[33] H.M. Hill, Physics Nobel honors foundational quantum entanglement experiments, Physics Today \textbf{75}, 14 (2022). 

\vspace{2mm}

[34] D.I. Kaiser, Tackling Loopholes in Experimental Tests of Bell’s Inequality, arXiv:2011.09296 (2020).

\vspace{2mm}
 
[35] J.S. Bell, \textit{Speakable and Unspeakable in Quantum Mechanics} (Cambridge U.P., Cambridge, 1987).

\vspace{2mm}

[36] P.C.W. Davies, J.R. Brown, \textit{El espíritu en el átomo: Una discusión sobre los misterios de la física cuántica} (Alianza, Madrid, 1989).

\vspace{2mm}

[37] A. Zeilinger, \textit{Dance of the Photons: From Einstein to Quantum Teleportation} (Farrar, Straus and Giroux, New York, 2010).

\vspace{2mm}

[38] J.L. Miller, Three groups close the loopholes in tests of Bell’s theorem, Physics Today \textbf{69}, 14 (2016). 

\vspace{2mm}

[39] B. Hensen, H. Bernien, A.E. Dréau, A. Reiserer, N. Kalb, M.S. Blok, J. Ruitenberg, R.F.L. Vermeulen, R.N. Schouten, C. Abellán, W. Amaya, V. Pruneri, M.W. Mitchell, M. Markham, D.J. Twitchen, D. Elkouss, S. Wehner, T.H. Taminiau, R. Hanson, Loophole-free Bell inequality violation using electron spins separated by 1.3 kilometres, Nature \textbf{526}, 682 (2015).

\vspace{2mm}
 
[40] J. Bernstein, \textit{Quantum Leaps}, (Belknap Press, Cambridge, Massachusetts, 2009).

\vspace{2mm}

[41] H.P. Stapp, Bell’s theorem and world process, Nuovo Cimento \textbf{29B}, 270 (1975) .

\vspace{2mm}

[42] A. Whitaker, \textit{The New Quantum Age: From Bell’s Theorem to Quantum Computation and Teleportation} (Oxford University Press, New York, 2012).

\vspace{2mm}

[43] E.J. Galvez, M. Beck, Quantum optics experiments with single photons for undergraduate laboratories, in: Tenth International Topical Meeting on Education and Training in Optics and Photonics  (SPIE, Ottawa, Ontario, 2007). 

\vspace{2mm}

[44] J.J. Thorn, M.S. Neel, V.W. Donato, G.S. Bergreen, R.E. Davies, M. Beck, Observing the quantum behavior of light in an undergraduate laboratory, Am. J. Phys. \textbf{72}, 1210 (2004). 

\vspace{2mm}

[45] D. Dehlinger, M.W. Mitchell, Entangled photons, nonlocality, and Bell inequalities in the undergraduate laboratory, Am. J. Phys. \textbf{70}, 903 (2002). 

\vspace{2mm}

[46] D. Dehlinger, M.W. Mitchell, Entangled photon apparatus for the undergraduate laboratory, Am. J. Phys. \textbf{70}, 898 (2002). 

\vspace{2mm}

[47] D. Branning, S. Bhandari, M. Beck, Low-cost coincidence-counting electronics for undergraduate quantum optics, Am. J. Phys. \textbf{77}, 667 (2009). 

\vspace{2mm}

[48] B.J. Pearson, D.P. Jackson, A hands-on introduction to single photons and quantum mechanics for undergraduates, Am. J. Phys. \textbf{78}, 471 (2010). 

\vspace{2mm}

[49] E.J. Galvez, Resource Letter SPE-1: Single-Photon Experiments in the Undergraduate Laboratory, Am. J. Phys. \textbf{82}, 1018 (2014). 

\end{document}